\documentclass[sigconf]{acmart}
\AtBeginDocument{%
  \providecommand\BibTeX{{%
    \normalfont B\kern-0.5em{\scshape i\kern-0.25em b}\kern-0.8em\TeX}}}

\setcopyright{rightsretained}
\copyrightyear{2023}
\acmYear{2023}
\acmDOI{10.1145/3586182.3615818}

\acmConference[UIST '23 Adjunct]{The 36th Annual ACM Symposium on User Interface Software and Technology}{October 29-November 1, 2023}{San Francisco, CA, USA}
\acmBooktitle{The 36th Annual ACM Symposium on User Interface Software and Technology (UIST '23 Adjunct), October 29-November 1, 2023, San Francisco, CA, USA}
\acmISBN{979-8-4007-0096-5/23/10}

\usepackage{lipsum}
\usepackage{caption}
\usepackage{subcaption}

\begin{document}

\title{FocusFlow: Leveraging Focal Depth for Gaze Interaction in Virtual Reality}


\author{Chenyang Zhang}
\authornote{Both authors contributed equally to this research.}
\email{zhang414@illinois.edu}
\orcid{0009-0003-1116-4895}
\affiliation{%
  \institution{University of Illinois at Urbana-Champaign}
  \streetaddress{201 North Goodwin Avenue MC 258}
  \city{Urbana}
  \state{Illinois}
  \country{USA}
}

\author{Tiansu Chen}
\authornotemark[1]
\email{tiansuc2@illinois.edu}
\orcid{0009-0007-7227-7948}
\affiliation{%
  \institution{University of Illinois at Urbana-Champaign}
  \streetaddress{201 North Goodwin Avenue MC 258}
  \city{Urbana}
  \state{Illinois}
  \country{USA}
}

\author{Rohan Nedungadi}
\email{rohanrn3@illinois.edu}
\orcid{0009-0001-8864-2467}
\affiliation{%
  \institution{University of Illinois at Urbana-Champaign}
  \streetaddress{201 North Goodwin Avenue MC 258}
  \city{Urbana}
  \state{Illinois}
  \country{USA}
}

\author{Eric Shaffer}
\email{shaffer1@illinois.edu}
\orcid{0009-0007-8656-7225}
\affiliation{%
  \institution{University of Illinois at Urbana-Champaign}
  \streetaddress{201 North Goodwin Avenue MC 258}
  \city{Urbana}
  \state{Illinois}
  \country{USA}
}

\author{Elahe Soltanaghaei}
\email{elahe@illinois.edu}
\orcid{0009-0006-5040-5438}
\affiliation{%
  \institution{University of Illinois at Urbana-Champaign}
  \streetaddress{201 North Goodwin Avenue MC 258}
  \city{Urbana}
  \state{Illinois}
  \country{USA}
}

\renewcommand{\shortauthors}{Chenyang, Tiansu, and Nedungadi, et al.}

\begin{abstract}
Current gaze input methods for VR headsets predominantly utilize the gaze ray as a pointing cursor, often neglecting depth information in it. This study introduces FocusFlow, a novel gaze interaction technique that integrates focal depth into gaze input dimensions, facilitating users to actively shift their focus along the depth dimension for interaction. A detection algorithm to identify the user's focal depth is developed. Based on this, a layer-based UI is proposed, which uses focal depth changes to enable layer switch operations, offering an intuitive hands-free selection method. We also designed visual cues to guide users to adjust focal depth accurately and get familiar with the interaction process. Preliminary evaluations demonstrate the system's usability, and several potential applications are discussed. Through FocusFlow, we aim to enrich the input dimensions of gaze interaction, achieving more intuitive and efficient human-computer interactions on headset devices.
\end{abstract}




\begin{teaserfigure}
 \centering
 \begin{subfigure}[b]{0.32\textwidth}
     \centering
     \includegraphics[width=\textwidth]{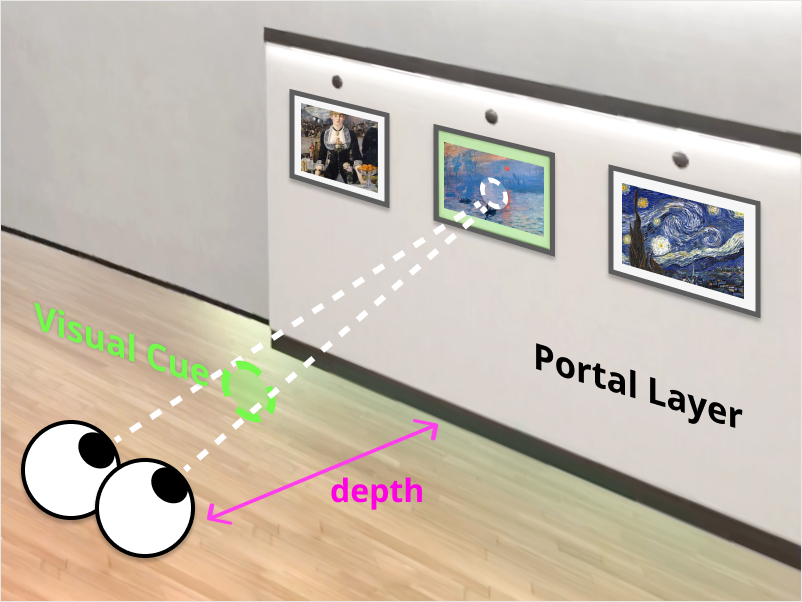}
     \caption{Hovering Over Target Object}
     \label{fig:teaser-b}
 \end{subfigure}
 \hfill
 \begin{subfigure}[b]{0.32\textwidth}
     \centering
     \includegraphics[width=\textwidth]{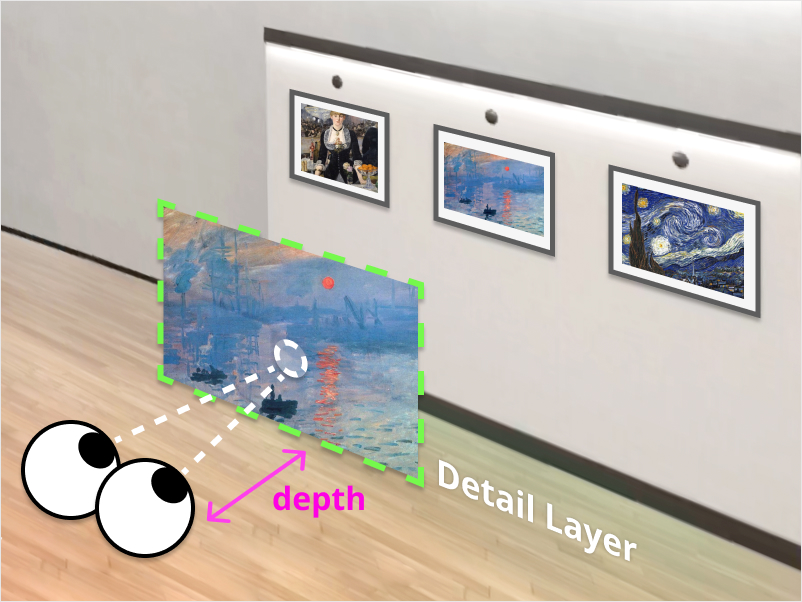}
     \caption{Activation Through Pulling In}
     \label{fig:teaser-c}
 \end{subfigure}
 \hfill
 \begin{subfigure}[b]{0.32\textwidth}
     \centering
     \includegraphics[width=\textwidth]{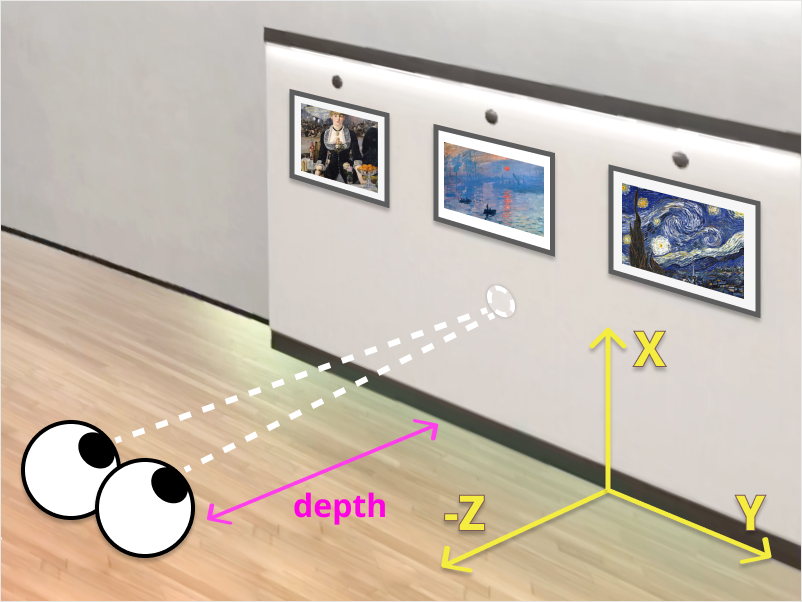}
     \caption{Exit Through Pushing Out}
     \label{fig:teaser-d}
 \end{subfigure}
\caption{Concept of FocusFlow. (a) In our approach, we calculate the focal depth using the parallax between two eyes. When users hover over an object, a visual cue will appear to guide the users to shift their focal point between layers in different depths. (b) Users can perform hands-free layer switch to activate a panel in front by shifting their focal depth. (c) Users can exit the detail layer by pushing out the focal depth along the z-dimension.}
\label{fig:teaser}
\Description{FocusFlow is a gaze interaction method that enhances traditional methods by incorporating the parallax of the user's eyes to calculate focal depth. By seamlessly integrating z-dimension information into the interaction logic, users can intuitively shift their focus between different layers, enabling hands-free selection.}
\end{teaserfigure}


\maketitle

\section{Introduction}

\par As humans, we naturally use our eyes to observe the environment and focus on objects. Taking advantage of this innate ability, a range of gaze interaction approaches have been explored for Virtual Reality (VR) and Augmented Reality (AR) headsets to enhance their input experience. Recently, Apple introduced Vision Pro, which first incorporates gaze input as a core component in commercial VR device interaction design. These gaze input methods typically utilize the gaze ray as a cursor for pointing at an object \cite{sidenmark2021radi, yi2022deep, ahn2021stickypie, yi2022gazedock, hedeshy2021hummer} and rely on other input behaviors such as hand gestures \cite{lu2021exploration, yu2021gaze} or dwelling \cite{lu2021evaluating, lu2021exploration, lu2021itext} to confirm selections.

\par However, unlike 2D screens where the cursor provides x-y position information, our focal point in the 3D world has an additional dimension along the z-axis. Despite this depth information being available in our sight, most studies and products only consider the gaze ray direction for pointing and neglect depth information. Although some works explore the application of binocular gaze information in the VR and AR space \cite{pai2016transparent, wang2022control}, none of them integrate the focal depth into current gaze interaction logic to propose a systematic interaction method. To fill this research gap, we propose FocusFlow, a novel gaze interaction method in VR that enables users to perform the layer switch by changing their focal depth with some visual guidance. Figure \ref{fig:teaser} illustrate how FocusFlow actively guides the users to shift their focal point between layers at different depths, enabling intuitive interaction with objects. This demo work contributes: 
(1) A sensitive and robust algorithm for detecting focal depth transitions by leveraging both eyes' gaze directions.
(2) A comprehensive interaction design for ``layer switch'' operation.
(3) Preliminary evaluations and applications demonstrating its usability and potential future usage.

\begin{figure}[htb]
  \centering
  \includegraphics[width=0.7\linewidth]{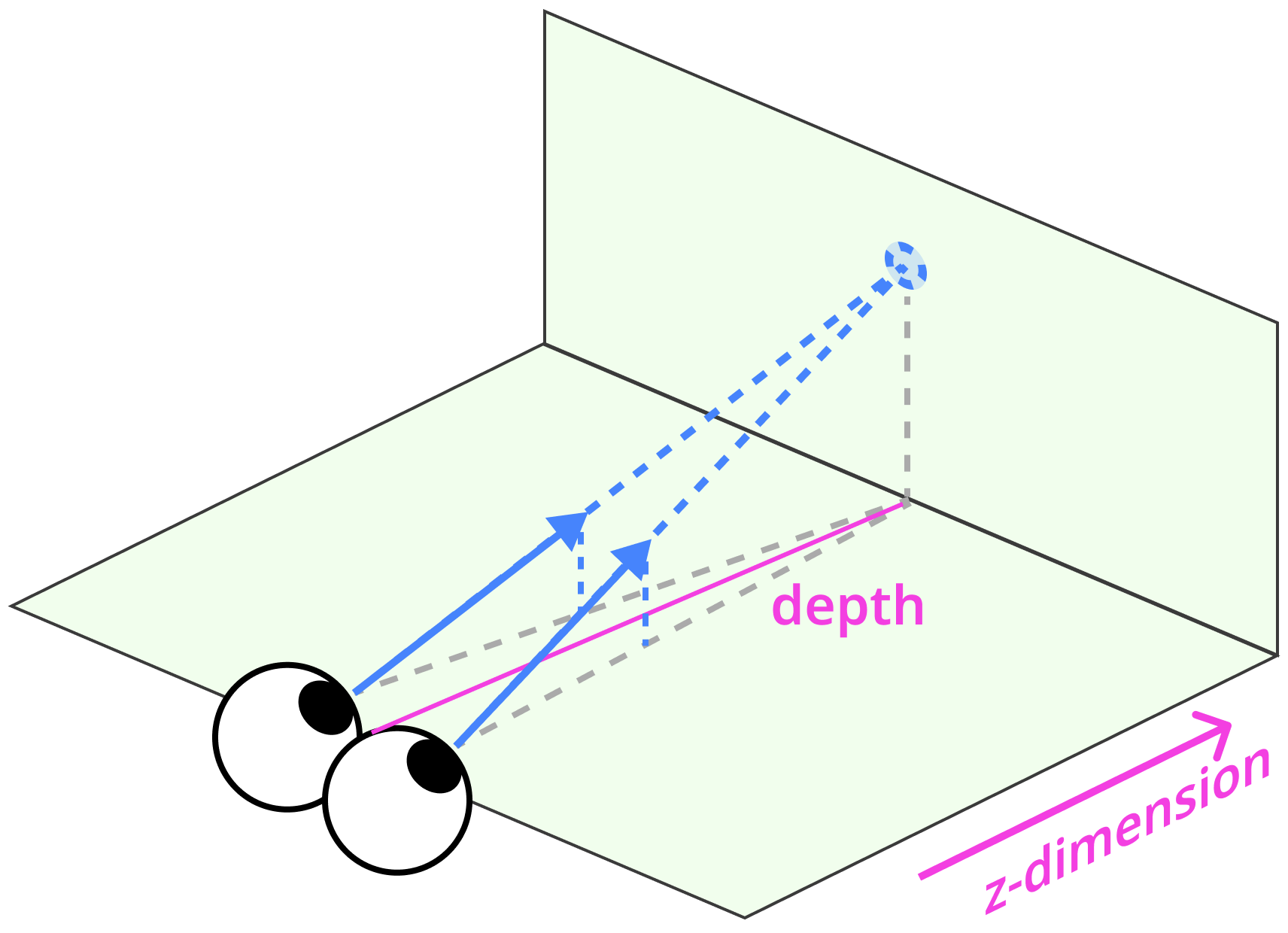}
  \caption{Depth Detection. Two gaze rays from both eyes intersect at a specific point, and the focal depth is the distance between the user and the intersection point along the z-dimension.}
  \label{fig:depthDetect}
\end{figure}

\section{FocusFlow}

\subsection{Detecting Focal Depth}

\par We determine focal depth by measuring the parallax between the two eyes. To achieve this, we utilize an eye-tracker built into the VR headset to capture the positions of the eyeballs. These positions are then converted into two gaze rays originating from both eyes and intersecting at a specific point (see Figure \ref{fig:depthDetect}). The focal depth is calculated as the distance between this intersection point and the user along the z-dimension. Additionally, since our eyes exhibit random movements when observing objects \cite{mughrabi2022my, zhang2021evaluating}, we have incorporated a de-noising algorithm to filter out random fluctuations in depth. For each time frame, we estimate the focal depth by current gaze directions and the depth data of recent history frames. This ensures that the resulting depth values are more stable and reliable.

\begin{figure}[htb]
  \centering
  \includegraphics[width=\linewidth]{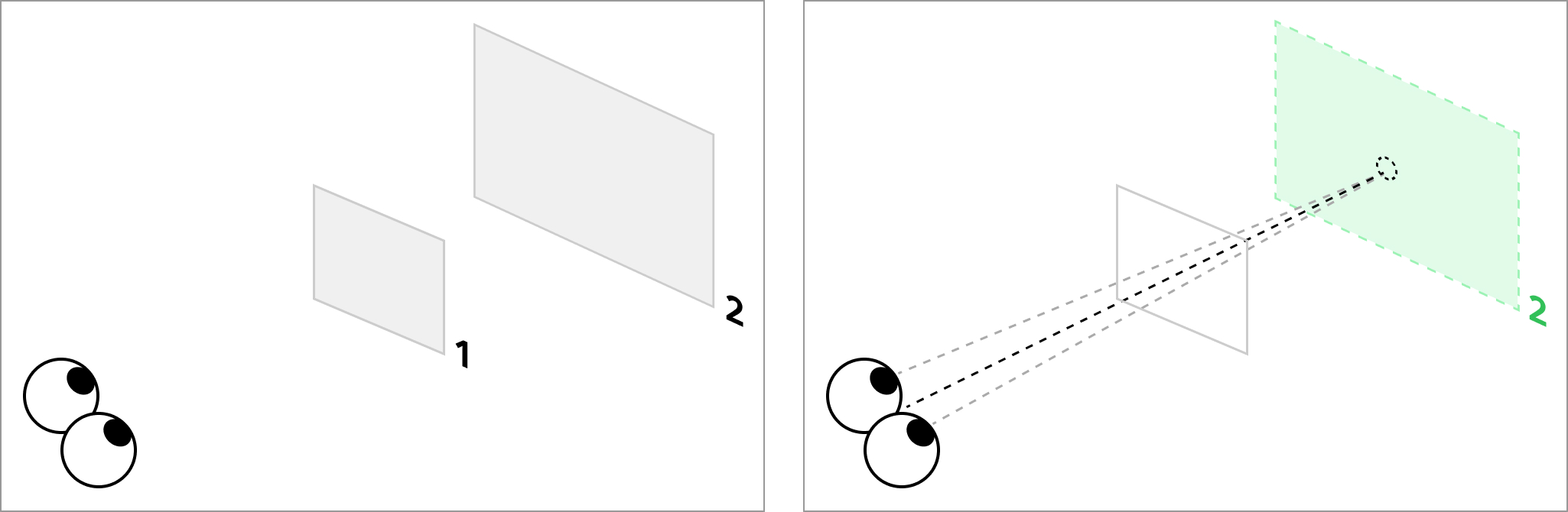}
  \caption{Layer-based UI and Depth Switch. Our proposed user interface consists of multiple layers with an information hierarchy. Users can freely switch between layers to get different levels of information by changing their focal depth.}
  \label{fig:layer}
\end{figure}

\begin{figure*}[htb]
  \centering
  \includegraphics[width=\linewidth]{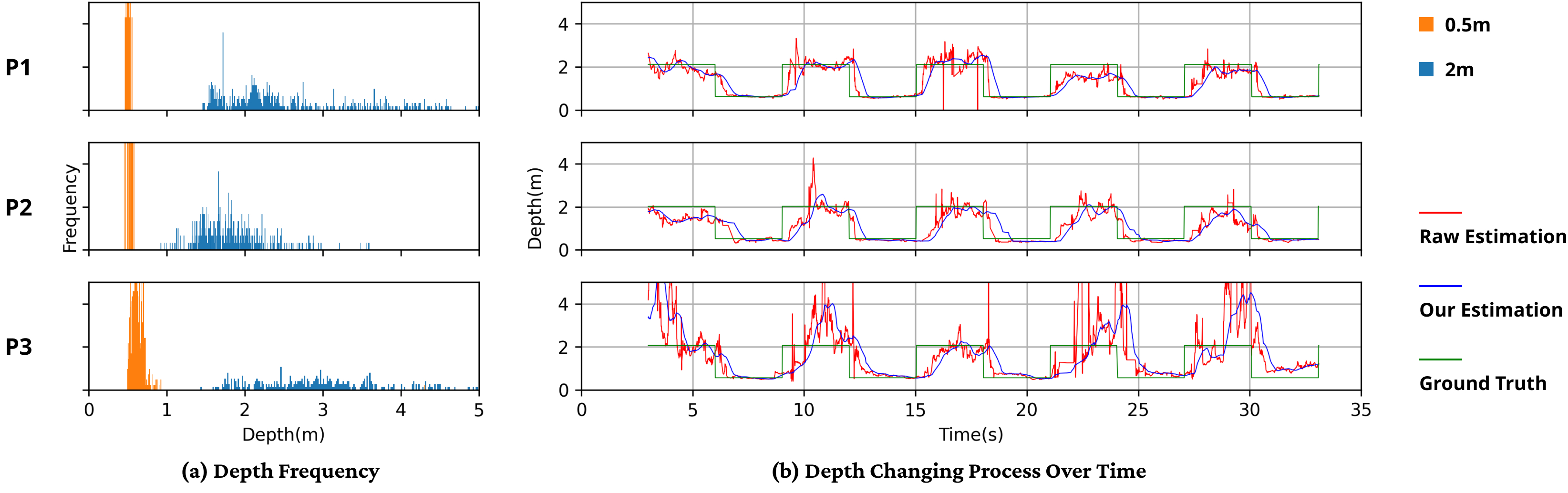}
  \caption{Preliminary evaluation result from three participants. (a) Participants are asked to stare at static targets at 0.5m and 2m respectively and their focal depth data are recorded. The depth frequency shows that our depth detection algorithm can distinguish between these two different depths. (b) Participants are then asked to follow a moving target between 0.5m and 2m in distance by their eyes. The depth curve illustrates our de-noising algorithm is capable of eliminating some outliers.}
  \label{fig:evaluation}
\end{figure*}

\subsection{Layer Switch through Depth Change}

\paragraph{\textbf{Layer-based UI}} Our focal depth changes as we observe objects at different distances. To naturally align this eye behavior with input process, we designed a layer-based User Interface (UI) that displays information on layers located at different distances along the z-dimension (Figure \ref{fig:layer}). Layers in different depths take advantage of the 3D space of the VR world and provide a new information placement space along the z-dimension. Through input in the z-dimension, users can choose any layer to view.
For the sake of simplicity, we set two layers in this demo: the objects in the VR world as the portal layer (Figure \ref{fig:teaser-b}), and a panel located near the user as the detail layer (Figure \ref{fig:teaser-c}). More layers can be set and customized to more usage scenarios.

\paragraph{\textbf{Depth Switch}} Users can activate different layers by changing their focal depth. In case of our example scenario, if users want to see detailed information about an object in the portal layer, they simply need to shift their focal point towards the detail layer closer to them (Figure \ref{fig:teaser-b} $\rightarrow$ \ref{fig:teaser-c}). When users need to return to the portal layer, they can just shift their focal point further to exit the activation (Figure \ref{fig:teaser-c} $\rightarrow$ \ref{fig:teaser-d}).  This provides users with an intuitive sense of ``grabbing in'' or ``taking a closer look'' at detailed information about selected targets. Similarly, shifting focal depth back towards the portal layer at the far end allows users to exit from viewing detailed information.

\subsection{Visual Cue}

\par It is worth noting that at the beginning of use, people may find it challenging to shift their focus onto something that is not visible. Therefore, it would be beneficial to offer users some cues that indicate the depth of hidden layers and assist them in adjusting the focal depth. To address this need, we incorporated a green circle in the center of the user's view as a visual cue. In the initial stages of use, users can rely on this visual cue positioned at their desired depth to quickly shift their focus and carry out interaction operations. As users become more acquainted with the system and develop muscle memory, we gradually remove these visual cues to minimize distractions and enhance immersion.

\section{Usability Analysis}

\par Two preliminary experiments were conducted to assess the usability and efficiency of FocusFlow. 
In the first experiment, participants were instructed to observe two objects in a VR scene: one located at a distance of 0.5 meters and another at a distance of 2 meters. We recorded the frequency distribution of the detected focal depth when observing the two objects as shown in Figure \ref{fig:evaluation}a. The results demonstrate that our detection algorithm is capable of distinguishing between two different focal depths, indicating that depth information can be collected as interaction input. However, it also reveals that for different participants, the visual depths obtained from the detection appeared to be shifted in various directions and degrees. This suggests that a personalized calibration scheme for depth detection algorithms is needed.

\par In the second experiment, we set up a scene where objects jump at distances of 0.5 meters and 2 meters to study the eye behavior during this transitioning process. Figure \ref{fig:evaluation}b illustrates how focal depth changes over time, suggesting that the eye movement is fast but with random noise present. The detected focal depth is accurate at close range, and the error becomes more significant at longer distances. Our de-noising algorithm effectively eliminates outliers and smooths the depth change curve; however, further improvement can be achieved by training additional machine learning models using eye datasets.

\section{Applications}

\par FocusFlow introduces a new dimension to interaction design, enabling the development of various applications. Here are some examples:

\begin{figure}[htb]
  \centering
  \includegraphics[width=\linewidth]{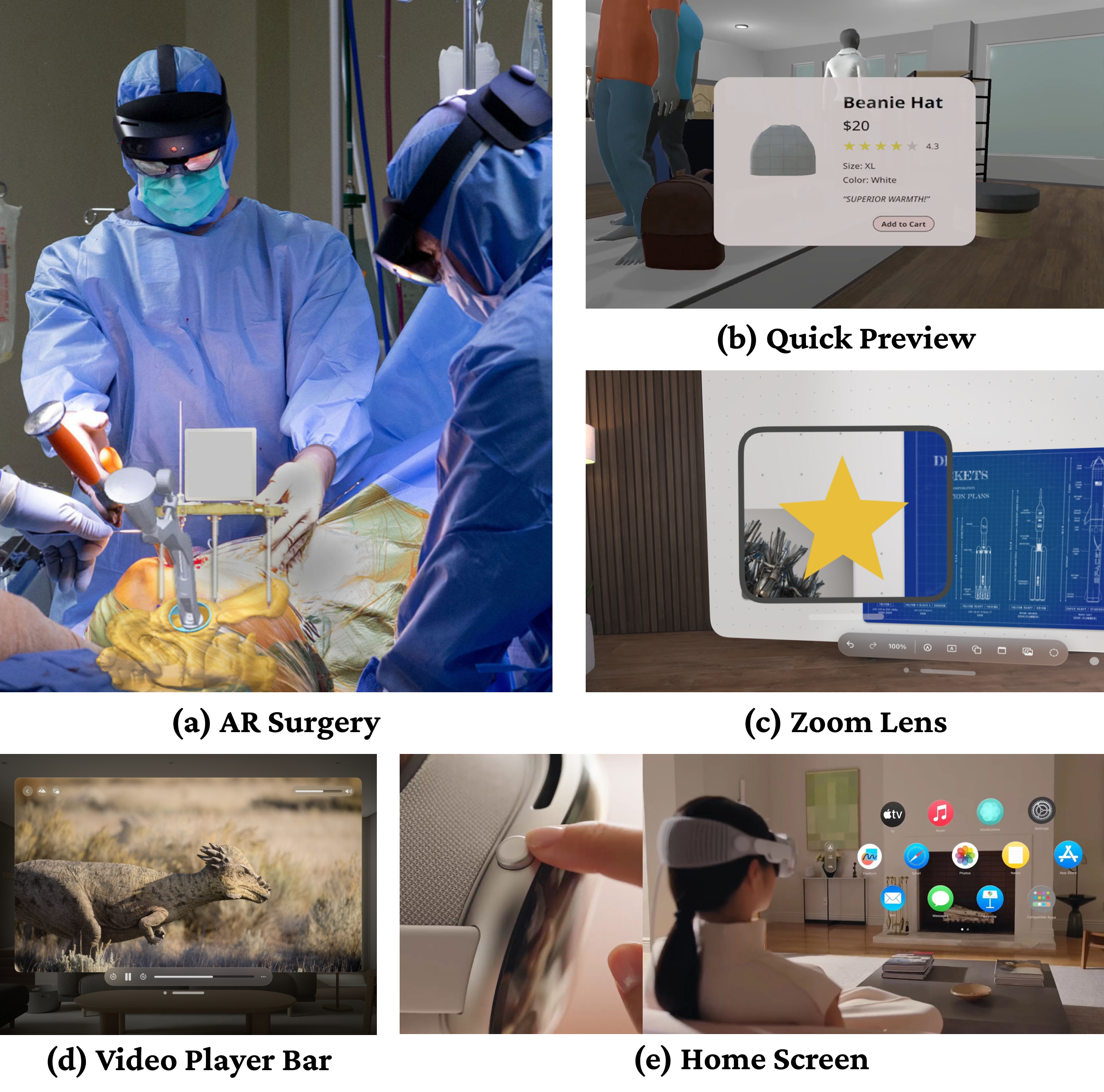}
  \caption{Possible applications for FocusFlow.}
  \label{fig:applications}
\end{figure}

\paragraph{\textbf{Hands-Free Selection}} In certain scenarios, using gestures for selection may not be an optimal solution. For instance, doctors performing surgery may have their hands occupied and cannot use gestures for quick interactions (Figure \ref{fig:applications}a). Similarly, lifting one's hand for gesture recognition during simple and quick preview interactions can be cumbersome and unnecessary (Figure \ref{fig:applications}b). Moreover, individuals with disabilities may be unable to use their hands for confirmations (Figure \ref{fig:applications}c). FocusFlow offers an intuitive and efficient solution by utilizing ``layer switch'' as the hands-free selection method.

\paragraph{\textbf{Activating Hidden Component}} Some UI components are hidden to enhance the immersive experience in VR environments. For example, when using certain applications, the home-screen is concealed, or when playing videos, the video player bar is hidden from view (Figure \ref{fig:applications}d). To activate these hidden components traditionally requires users to perform specific gestures or press designated buttons which disrupts their immersion in the virtual environment (Figure \ref{fig:applications}e). With FocusFlow, users can naturally activate these hidden components simply by shifting their attention between different depths, which is intuitive and more immersive.

\begin{figure}[htb]
  \centering
  \includegraphics[width=\linewidth]{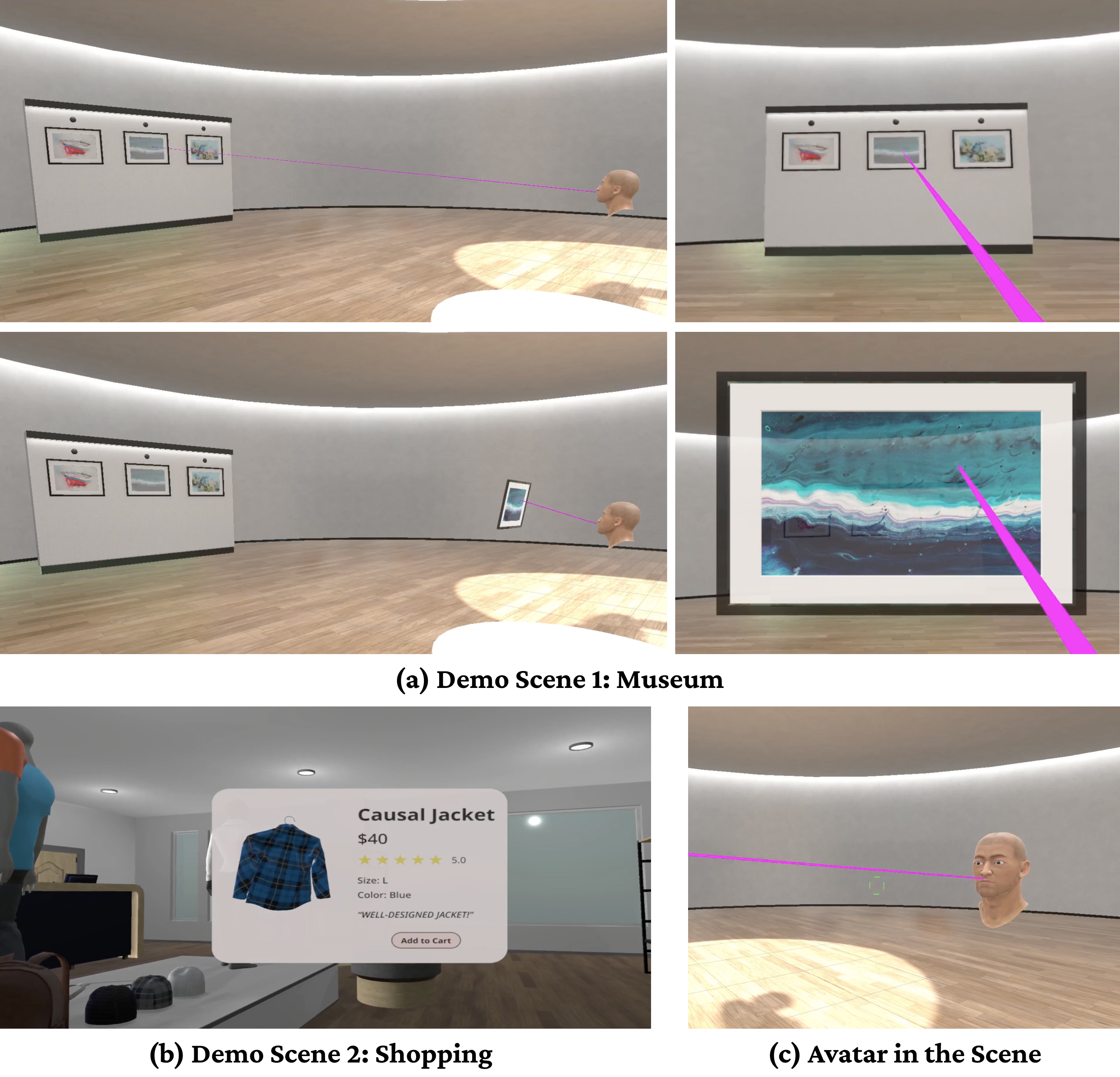}
  \caption{Demonstration Scene in Unity.}
  \label{fig:scene}
\end{figure}

\section{Demonstration plan}

In the demo, we will demonstrate the FocusFlow in two VR scenes, the art museum (Figure \ref{fig:scene}a) and the fashion store (Figure \ref{fig:scene}b). We adopt HTC Vive Pro Eye as our VR hardware, which enables the eye-tracking feature. Users will have an immersive experience of the FocusFlow when performing the layer switch by actively changing their focal depth.


\bibliographystyle{ACM-Reference-Format}
\bibliography{sample-base}











\end{document}